\documentclass[fleqn,10pt]{wlscirep}

\usepackage{times}
\usepackage[version=1]{mhchem}
\usepackage{graphicx}
\usepackage{graphics}
\usepackage{amsmath}
\usepackage{color}
\usepackage{amssymb}
\usepackage{algorithmic}
\usepackage{array}
\usepackage{float}
\usepackage{soul}
\usepackage{cite}
\usepackage{url}
\usepackage{setspace}
\DeclareGraphicsExtensions{.jpg,.pdf}

\title{Enhancing Intrinsic Stability of Hybrid Perovskite Solar Cell by Strong, yet Balanced, Electronic Coupling}

\author[1]{Fedwa El-Mellouhi}
\author[1]{El Tayeb Bentria}
\author[1]{Sergey N Rashkeev}
\author[1,2,3]{Sabre Kais}
\author[1,2,*]{Fahhad H Alharbi}
\affil[1]{Qatar Environment and Energy research Institute (QEERI), Hamad Bin Khalifa University, Doha, Qatar}
\affil[2]{College of Science and Engineering, Hamad Bin Khalifa University, Doha, Qatar}
\affil[3]{Department of Chemistry, Physics, and Birck Nanotechnology Center, Purdue University, West Lafayette, Indiana 47907, USA}
\affil[*]{falharbi@qf.org.qa}

\begin{abstract}
In the past few years, the meteoric development of hybrid organic--inorganic perovskite solar cells (PSC) astonished the community. The efficiency has already reached the level needed for commercialization; however, the instability hinders its deployment on the market. Here, we report a mechanism to chemically stabilize PSC absorbers. We propose to replace the widely used methylammonium cation (\ce{CH3NH3+}) by alternative molecular cations allowing an enhanced electronic coupling between the cation and the \ce{PbI6} octahedra while maintaining the band gap energy within the suitable range for solar cells. The mechanism exploits establishing a balance between the electronegativity of the materials' constituents and the resulting ionic electrostatic interactions. The calculations demonstrate the concept of enhancing the electronic coupling, and hence the stability, by exploring the stabilizing features of \ce{CH3PH3+}, \ce{CH3SH2+}, and \ce{SH3+} cations, among several other possible candidates. Chemical stability enhancement hence results from a strong, yet balanced, electronic coupling between the cation and the halides in the octahedron. This shall unlock the hindering instability problem for PSCs and allow them to hit the market as a serious low-cost competitor to silicon based solar cell technologies. 
\end{abstract}
\begin{document}

\flushbottom
\maketitle

\thispagestyle{empty}

\section*{Introduction}

In the past few years, the solar cell community has witnessed an exceptional emergence of a new family of solar cell materials\cite{K01,S01,B01}; namely hybrid perovskite solar cells (PSC). Within just four years, the conversion efficiency has ramped up dramatically and reached 20.1\%\cite{Y01}. This dramatic development is believed to be a result of a unique supportive combination of different properties of these materials, including the favorable balance between strong absorption and long carrier lifetime\cite{M01}, the efficient transport\cite{R01,P01,D01}, and the benign fault tolerance\cite{G01}. From a practical perspective, it is also impressive how simple to fabricate the PSCs and how many efficient cells made with various hybrid perovskites absorbers (the mostly used compound of this family is \ce{CH3NH3PbI3}) and with different device designs. The only remaining obstacle before large scale commercialization is the cells instability\cite{K01,S01,B01}.

\begin{figure*} [t]
\centering
\includegraphics[width=6.0in]{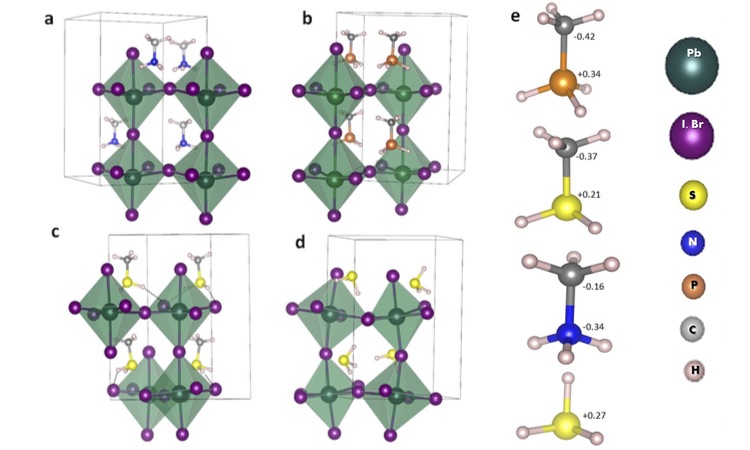}
\caption{Crystal structures of the tetragonal phases of (\textbf{a}) \ce{CH3NH3PbX3}, (\textbf{b}) \ce{CH3PH3PbX3}, (\textbf{c}) \ce{CH3SH2PbX3}, and (\textbf{d}) \ce{SH3PbX3} where \ce{X}=\ce{I} or \ce{Br}. (\textbf{e}) The partial charges of the used cations \cite{N02}.} 
\label{Str}
\end{figure*} 

Currently, it is well known that \ce{CH3NH3PbI3} is not stable; this is due to many extrinsic and intrinsic causes. Extrinsically, it is sensitive to moisture, UV exposure, and oxygen\cite{B01,Z01}. Another important aspect that contributes significantly to the instability is the moderate crystal quality\cite{S03}, that ignites consequently an additional mixture of instability issues. Moreover, there is a debate about the severity of the dynamics of the polarized molecular cations such as \ce{CH3NH3+} \cite{M04,C01}, beleived now to contribute to many materials related characters. More fundamentally, \ce{CH3NH3PbI3} suffers from intrinsic instability that can result in disorder and hence larger defect density, assist phase transition, and make the materials thermally active. Recently, Zhang et al.\cite{Z01} confirm computationally that the compound is thermodynamically unstable and suggested  the existence of a kinetic barrier that prevents its spontaneous decomposition to \ce{CH3NH3I} and \ce{PbI2}; however, the decomposition is inevitable in the long-term.

Structure-wise, the 3-dimensional (3D) \ce{CH3NH3PbI3} is composed of a  network of corner-sharing \ce{PbI6} octahedra and molecular cations (\ce{CH3NH3^+}) hosted between the cages. The main features of the resulting electronic structure are: 1) the top of valance band is composed of the $5p$ orbitals of the iodine, 2) the edge of the conduction band is formed from the $6p$ orbitals of the lead, and 3) the electronic states due to \ce{CH3NH3+} are located several electonvolts above and below the band gap edges and they don't contribute directly neither to the optical properties within the solar spectrum range nor to the electronic transport\cite{M01,F01}. These facts entail many other consequences: first, the cation can be used ``indirectly'' to tune the optical and electrical properties by distorting the octahedral network. Filip et al.\cite{F01} and Knutson et al.\cite{K02} utilized this concept to tune the gap in 3D and 2D hybrid perovskites respectively. Secondly, the cohesion within the crystal between \ce{PbI6} octahedra and \ce{CH3NH3^+} is mainly due to weak electrostatic interactions\cite{W01} as the electronic coupling between the octahedron and the cation is negligible. Thus, it was found that the cohesion is relatively weak\cite{B01,W01} as characterized by the relatively small site Madelung potential\cite{W01,F02} leading to  chemical instability.

Currently, molecular cation design and substitution approach is an active area of research where the focus has been  almost utterly on tuning the optoelectronic properties\cite{F01,F02,A01}. While, this could help in further enhancing the efficiency of PSCs; their commercial deployement is on hold till instability issues are resolved. Here, we demonstrate that the cation design can be used as a mechanism to enhance the stability by triggering stronger electronic coupling and electrostatic interactions like hydrogen-bonding, halogen-bonding, and van der Waals. This constitutes a  revival of  mechanisms  routinely utilized other disciplines such as  for the construction of polymers and metal Organic frameworks\cite{D02,G03,S04,L02} to the world of PSC.

\begin{figure*} [t]
\centering
\includegraphics[width=4.5in]{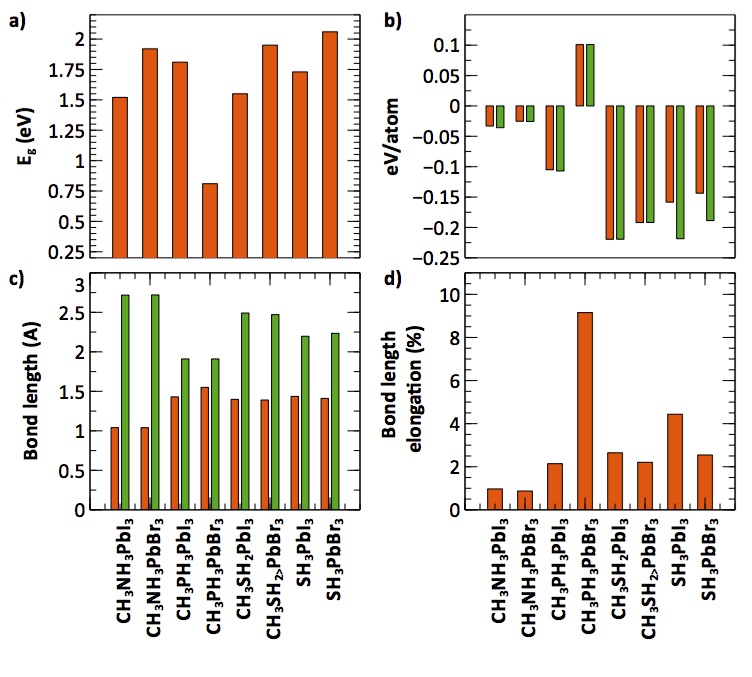}
\caption{(\textbf{a}) The band gap energy of the studied materials. (\textbf{b}) The calculated reaction (red) and hull (green) energies (in eV/atom). (\textbf{c}) The bond lengths for the bridging \ce{H} atom and both its cation donating atom (red) and the halide (green). (\textbf{d}) The bond length elongation between the bridging \ce{H} and its atom (\ce{N}, \ce{P}, or \ce{S}) in the cation in the crystal compared to its length as a stand alone cation\cite{N02}.}
\label{Fig01}
\end{figure*} 
\begin{figure*} [t]
\centering
\includegraphics[width=6.0in]{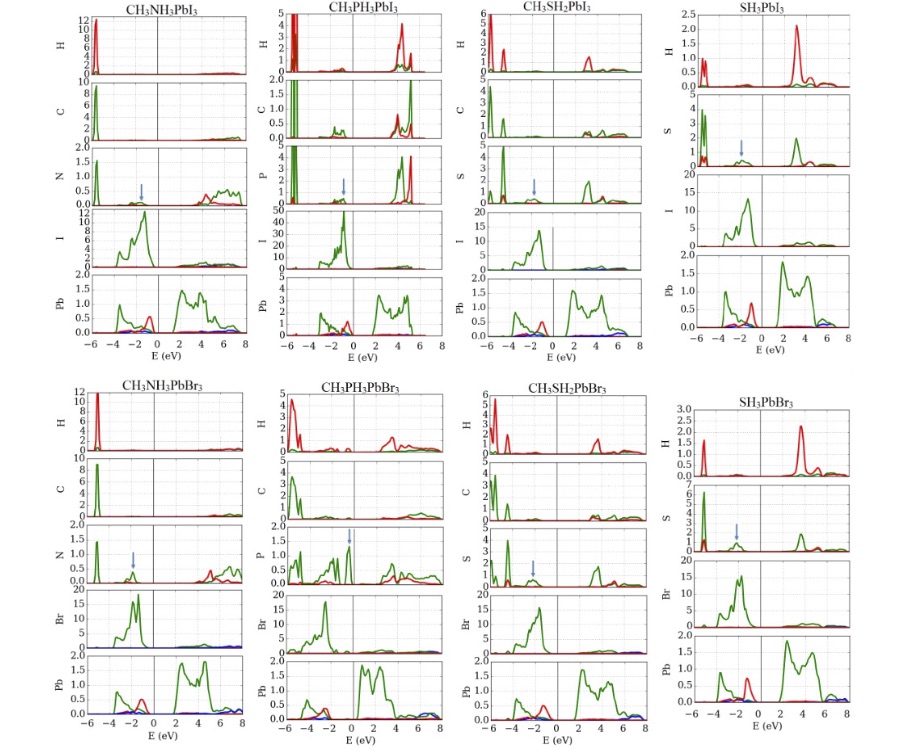}
\caption{The projected density of states (PDOS) of all the studied materials (red for $s$ orbitals, green for $p$ orbitals, and blue for $d$ orbitals). The arrows indicate the bridging states for which we calculate normalized PR values.}
\label{PDOSFig}
\end{figure*}

In this work, we investigated computationally using density functional theory (DFT), the possibility of manipulating the electronic coupling between the molecular cation and \ce{PbI6} octohedron to enhance the stability of hybrid perovskites materials. The commonly used \ce{CH3NH3+} cation is composed of the strong electronegative atom of \ce{N} and the moderate electronegative atom of \ce{C}. By considering the electronegativity of \ce{I}, the hydrogen atoms are generally tightly bonded to \ce{CH3NH3+} making its interaction with the \ce{PbI6} octahedron fairly small. By replacing the \ce{N} atom by a less electronegative atom, the binding  of the hydrogen atoms to the cation can be reduced offering the possibility to enhance the electronic coupling with \ce{PbI6} octahedron. We found that this indeed improves the chemical stability considerably, which is quantified by the reaction\cite{H02} and the hull\cite{C03} energies. The reaction energy is the difference between the total energy of a reaction and reactants, whereas the hull energy is the difference in formation energies and it effectively evaluates the stability of a given compound against any linear combination. While many possibilities for phase separation exist;  we used the separation into the most stable binary and ternary solid-state compounds based on phase diagrams contructed from the Materials' Project database \cite{J01}. 

For sake of systematic analysis, we focused only of the tetragonal phases at which \ce{CH3NH3PbI3} crystallizes at room temperature. Our DFT stability calculations show that both \ce{CH3NH3PbI3} and \ce{CH3NH3PbBr3} are  marginally stable in agreement with previous studies\cite{Z01,K04}. By replacing the molecular cation with the purpose of enhancing the electronic coupling between the cations and \ce{PbI6} octahedra, we found that \ce{CH3PH3+}, \ce{CH3SH2+}, and \ce{SH3+} cations result in more stable hybrid perovskite crystals while maintaining the suitable energy gap for solar cells. The only exception is \ce{CH3PH3PbBr3}, where the interaction (P-H-Br) becomes extremely strong causing the bridging bond (P-H) elongation by 10\% and unbalancing of the interactions and chemical destabilization of  material. The strength of the electronic coupling is accessed from electronic structure and electronic density analyses,  measures of relevant bond elongation, and electronic delocalization given  in terms of computed normalized participation ratio (PR).

\section*{Results \& Discussion}
\begin{figure*} [t]
\centering
\includegraphics[width=6.0in]{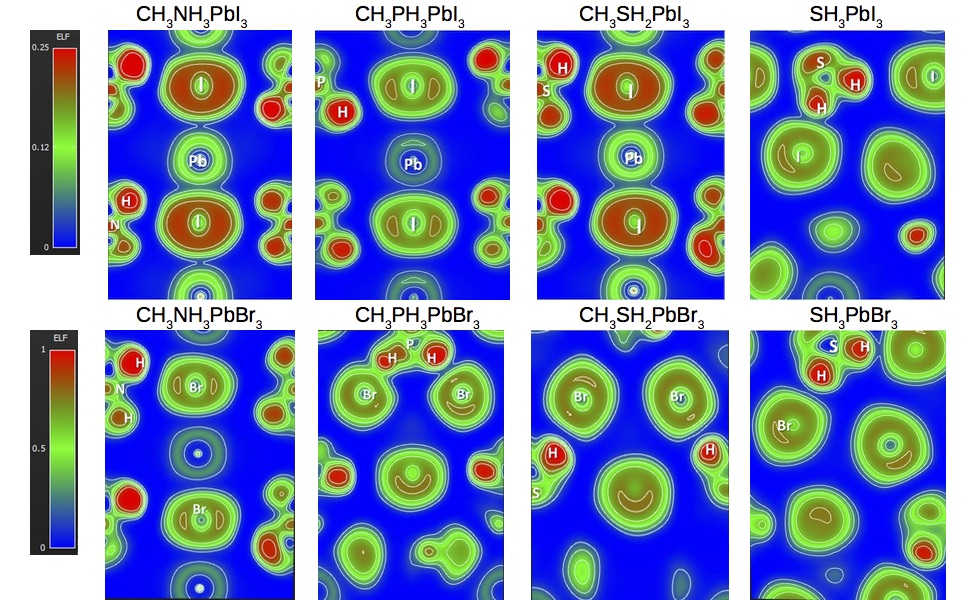}
\caption{The contour maps of the electronic densities of the bridging states at planes with maximum interaction.}
\label{FigED}
\end{figure*}

As we aim to replace the methylammonium  cation in \ce{CH3NH3PbI3} and \ce{CH3NH3PbBr3} to improve the stability without deteriorating the suitable energy gap ($E_g$), the cation must be relatively small to maintain the 3D dimensionality of the hybrid perovskites and must not contain any of the high electronegative elements to allow dragging (binding) one of its \ce{H} atoms more toward the \ce{PbI6} or \ce{PbBr6} octahedron. There are many possible cations satisfying this condition and not limited to the ones considered in this work, namely \ce{CH3PH3+}, \ce{CH3SH2+}, and \ce{SH3+}. The relaxed crystal structures in tetragonal phases with the new cations are shown in Fig.-\ref{Str}, for which, the reaction and hull energies are calculated and plotted in Fig.-\ref{Fig01} besides the band gap energy ($E_g$) and the relevant bond elongation. Here, we present the inverse energy to the hull, hence  the more stable compounds pocess  the  most negative energies. The relevant bridging bond is between the  electron donating element  in cation (P, N, S) and its farthest \ce{H} atom forming a bridge with the halide atom (I, Br) are reported.

Clearly, the band gap (see Fig.-\ref{Fig01}\textbf{a}) is slightly affected by cation substitution, but remains suitable for solar cell applications except for \ce{CH3PH3PbBr3}, where it is reduced considerably due to the emergence of an intermediate state resulting from a formed bond between \ce{Br}, the bridging \ce{H} atom, and \ce{P} \cite{F03}. As will be shown shortly, this results in losing the electrostatic balance and hence deteriorating the electronic coupling and chemical stability.

The most significant ramification of the enhanced electronic coupling affects the stability; the stable reaction and hull energies (Fig.-\ref{Fig01}\textbf{b}) are considerably improved by a factor of --at least-- 4 compared to \ce{CH3NH3PbI3} and \ce{CH3NH3PbBr3}, except for \ce{CH3PH3PbBr3} which is completely unstable. As can be seen, both \ce{CH3NH3PbI3} and \ce{CH3NH3PbBr3} are  marginally stable with reaction energies ranged between -25 and -30 meV/atom. In such case, the thermal energy at moderate temperature (50-100$^{\circ}$C) is enough to decompose the materials. Such decomposition does not happen spontaneously due to some kinetic barrier \cite{Z01}. For all the other materials, the calculated reaction and hull energies stand between -230 and -100 meV. Worth-notingly, this behavior is correlated with the elongation of the bridging bond between the \ce{H} atom and its donating atom (Fig.-\ref{Fig01}\textbf{d}) in the cation compared to the standalone molecule\cite{N02}. In the cases of \ce{CH3NH3PbI3} and \ce{CH3NH3PbBr3}, the bonds are barely stretched by 0.97\% and 0.87\% respectively. This is mainly due to the high electronegativity of the \ce{N} atom, which keeps the bonded \ce{H} atoms strongly gripped. In term of normalized PR (Table-\ref{TB1}), the contributions from both of them to the bridging states (indicated by arrows in Fig.-\ref{PDOSFig}) is small, implying that the associated \ce{H} atom is highly localized within the cation and almost not interacting with the octahedron. On the other hand, the bridging \ce{P}-\ce{H} and \ce{S}-\ce{H} bonds are elongated much further and their corresponding normalized PRs are --at least-- doubled.
For the stable materials, the elongation (Fig.-\ref{Fig01}\textbf{d}) ranged between 2.1\% and 4.4\%, which is mainly due to the attraction of \ce{H} atom to the halide and the bridge formation. So, the electronic densities of \ce{P}, \ce{S}, and \ce{H} are delocalized due to the elongation and hence their normalized PRs are increased. This attraction becomes even more substantial in \ce{CH3PH3PbBr3} where the bond is 9.1\% stretched. In this particular case, the bridging \ce{H} atom is attracted significantly by the halide till breaks the balance around it. We attribute this phenomena to the electronegativity difference between \ce{Br} (2.96) and \ce{P} (2.19) \cite{H03} among others such as partial charges and polarizability (Fig.-\ref{Str}\textbf{e}). 

\begin{table*} [t]
\small
\centering
\caption{The Normalized participation ratios for the bridging states (Fig.\ref{PDOSFig}) between the cations and the octahedra.}
\begin{tabular}{|l|c|c|c|}
\hline 
 & \ce{N}, \ce{P}, or \ce{S} & \ce{H} & \ce{I} or \ce{Br}  \\
\hline
\ce{CH3NH3PbI3} & 0.018 & 0.009 & 0.973 \\
\hline
\ce{CH3NH3PbBr3} & 0.049 & 0.005 & 0.946 \\
\hline
\ce{CH3PH3PbI3} & 0.068 & 0.034 & 0.898 \\
\hline
\ce{CH3PH3PbBr3} & 0.694 & 0.056 & 0.250 \\
\hline
\ce{CH3SH2PbI3} & 0.073 & 0.018 & 0.909 \\
\hline
\ce{CH3SH2PbBr3} & 0.076 & 0.017 & 0.907 \\
\hline
\ce{SH3PbI3} & 0.124 & 0.013 & 0.863 \\
\hline
\ce{SH3PbBr3} & 0.202 & 0.011 & 0.787  \\
\hline
\end{tabular} 
\label{TB1}
\end{table*}

So far, we mainly discussed the structural effects of the enhanced electronic coupling between the cation and the octahedron. Further understanding requires a close look at the electronic structure. Fig.-\ref{PDOSFig} show the projected density of states (PDOS). For \ce{CH3NH3PbI3} and \ce{CH3NH3PbBr3}, the contributions of \ce{CH3NH3+} are deep in the valence and conduction bands. There are barely no signs for interactions between the cation and \ce{PbI6} and \ce{PbBr6} octahedra. This is severely altered by replacing by the suggested cations. Several molecular states caused by the enhanced electronic coupling appear at the top of the valance band. In the extreme case of very strong interaction, it results in emerging states occupying the top of the valance band and considerably shifting  its edge as in the case of \ce{CH3PH3PbBr3} (Fig.-\ref{PDOSFig}).

To visualize the electronic localization, the contour maps of the electronic densities of the bridging states at the planes with maximum interaction are plotted and shown in Fig.\ref{FigED}. In the case of \ce{CH3PH3PbBr3}, it is clear that a very strong and connected bridge is formed as typical for hydrogen bonding. For \ce{SH3PbI3} and \ce{SH3PbBr3}, the couplings between \ce{SH3+} cation and the halides are reasonably strong; but not to the level needed to form hydrogen-bond.

\section*{Conclusion}
In conclusion, we show how to chemically stabilize the hybrid perovskite solar cell absorbers by replacing methylammonium  cation (\ce{CH3NH3+}) by other cations that enhance the electronic coupling between the molecule and the octahedra while maintaining the band gap energy within the suitable range for solar cells. Practically, this is attainable by exploiting the electronegativity of the materials' constituents and the resulting electrostatic interactions. Our calculations show that the stability is correlated to balancing the interaction and hence the electronic coupling between the halides in the octahedron and the cations. In this work, we considered \ce{CH3PH3+}, \ce{CH3SH2+}, and \ce{SH3+} cations where the reaction and hull energies are enhanced by a factor of --at least-- 4; however, several other molecular cations can be used where the coupling and hence the stability can be enhanced.

\section*{Methods}
We employ density functional theory (DFT) calculations to evaluate the electronic structure and estimate the stability of the proposed materials. The DFT calculations are performed with the projector augmented wave (PAW) method as implemented in the Vienna Ab-initio Simulation Package (VASP) \cite{K03}. For the exchange correlation energy of interacting electrons, the generalized gradient approximation (GGA) with the parameterization of Perdew--Burke--Ernzerhof (PBE) \cite{P02} is used. The energy cutoff for the planewave basis set was set to 520 eV and an $8 \times 8 \times 8$ Monkhorst--Pack $k$-point mesh is employed where the convergence on the final forces is set at 0.01 eV/\AA. For the stability calculations, the phase diagrams (not shown due to space limit) were generated using PyMatGen with the Material project (MP) DataBase \cite{J01}. The convex-hull \cite{C03} construction method effectively evaluates the stability of a given compound against any linear combination. This value is obtained by means of DFT and Phase Diagrams calculations. PR is obtained directly from VASP's output for the relaxed structures.

\bibliographystyle{naturemag}
\bibliography{PSCS}

\section*{Acknowledgements}

We are extremely grateful for the Research Computing Center in Texas A\&M University at Qatar and SHAHEEN Supercomputer at King Abdullah University of Science and Technology (KAUST), Suadi Arabia, where the calculations were conducted.

\end{document}